# A Principle of Regulating the Collective Effect of Assembling Patterns in a Moderate Number of Equivalent Finite Regular Arrays of Active Nanoelements due to Local Transferring Information by Pairs Hopping under Synchronous Excitation

Wlodzimierz Kozlowski*

*Inst. of Biocybernetics & Biomed. Eng. PAS, Trojdena 4, 02-109 Warsaw, Poland,
wlodekak@ibb.waw.pl , wlodekak@icm.edu.pl , wlodekak@fastmail.fm

## ABSTRACT

One summarizes finding of our recent computational research with a prospect of relevant feasibility study of regulating the process of assembling nano-sized elements into functional materials and systems. This shows that accomplishing the collective effect (CE) of assembling the active nano-elements into functional quasi two dimensional patterns over finite surfices within a number $M$ of subsystems, being of the same type but different in details, may be regulated by changing the ambient conditions. This regulation of the CE requires that the immobile nano-elements, arranged into regular array over the surface, are to be activated due to local information transfer realized by hopping nano-sized elements, each represented by a pair of its ends only: sending end and the receiving one; two neighbor sites of the array identify position of that pair. Then, the CE may be regulated by varying the $M$, a common forcing factor and efficiency of projecting the regularity of the array onto the CE.

***Keywords***: collective effects, regulation, finite size effects, discrete systems, prospect of feasibility

## 1  THE NEED

One may note the emerging task of regulating the low cost process of self-assembling the nanometer size elements into functional systems such as information processing materials, sensors, shells and covers. The life sciences give evidences that living nature provides examples of organizing various nano-sized elements into sufficiently stable functional structures by making use of surfaces and water molecules utilizing their specific structural and physico-chemical features (this is very strong simplification, of course). However, if our intention is to extract such benefits from arranging the nano-sized elements into functional systems and their further transformations that, in principle, cannot result from the known natural processes of life or if the relevant processes of life would appear too complex for controlable way of utilizing them, we will have to design artificial methods of *spatial* regulating the processes of assembling the nano-sized elements into functional systems and *spatial* control of processing information by these systems.

This emerging task reveals the need in indicating general requirements, fulfilling of which would allow solving it. With the purpose to indicate these general requirements the reported computational study has been performed. Here, we follow the way of presentation using more general terms than in our extensive report [1] to point out only the elements of our approach which enable us to reveal the prospect of feasibility study.

## 2  SOLUTION: THE MODEL

### 2.1  Assumptions

We search for methods of *regulating* the collective outcome of assembly processes taking place in an ensemble of a moderate number $M$ of finite complex systems, of the same type but different in details, evolving simultaneously and independently, under influence of stochastic factors within equivalent finite spatial domains and in the same ambient conditions. The finite discrete spatial domains are equivalent regular arrays in all the component systems. Information transmission processes within these domains result in manifestations represented by various patterns of expanding spatial regions specified with the same feature in all the systems at each stage of the process. One assumes that these regions specified within all the component systems affect simultaneously and independently certain receiver system in the equivalent conditions. Moreover, differences between contributions, which distinct component systems have to the effect of their influence on the receiver system, are attributed only to differences between spatial arrangements of the regions specified within these component systems. Thus the receiver system recognizes collective outcome of evolution processes within all the component systems. The result of this influence manifests itself in the receiver system as a **collective effect - CE** that can be represented at each stage of the process by certain pattern of distinguished regions within a spatial domain being equivalent to the domain of a component system. Accordingly, changing certain values parameterizing the ambient conditions of evolution of all the component systems appears to be a way of regulating the CE accomplishment.

### 2.2  Model Scenario

Certain complex molecules or nano-structured molecular complexes are thought so immobilized at the surface that they are arranged into the finite regular hexagonal array and their ability to be activated or deactivated is their only recognizable feature; we call them **site elements**. Appearance of that model feature at a site (i.e., activation of a site element) is indicated by covering this site and thus the pattern of sites covered represents a state of the component system at the stage of evolution.

One investigates random expansion process (REP) as Markov process of covering sites of finite regular hexagonal array. States of the REP at stages $T$ are finite random sets (FRS), $F(T) = \{A_i(T), i = 1, 2, ..., M\}$ with realizations $A_i(T)$ developing simultaneously and independently in the same model ambient conditions. Accomplishing of the model collective effect (CE) is represented by sequence of patterns, denoted as $\int \mathbf{F}(\mathbf{T})$, each of which is mean expected form of all the $A_i(T)$ and computed as **mean measure set - MMS** after known algorithm [2] (see also Appendix in [1]).

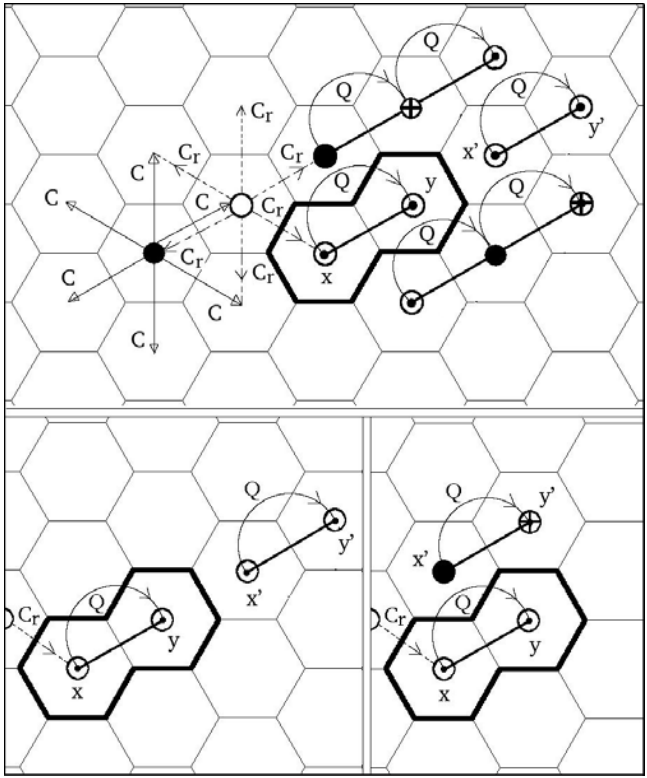

Figure 1: Effective modeling a path of the local information transfer in a component system $A_i$ for one step of the evolution, from $T$ to $T+1$. (hopping from only one initial position of the nanowire is depicted); Here: ● - sites covered at stage $T$, ⊕ - sites that can be covered for the $T+1$, ○ - sites not covered either at $T$ or $T+1$, (• within ○ or ⊕ indicate ends of a nanowire).

In each realization $A_i$ evolving from a stage $T$ to $T+1$, the information transfer between sites of the array is assumed as being realized by finite size, effectively one dimensional nanoelements transferring finite information portions, called signals, while hopping to positions close by; they are identified here as the **model nanowires** because they can also transmit a signal between their ends. Specifically, transmission of the signal changes binding properties at the nanowire ends and this can affect behavior of the site elements occupying sites coinciding with these ends. For modeling effects of this information transfer, we generalize our approach used previously [3] for modeling the contribution which certain finite-size elements representing organized fluid streams and structures have to the collective effect of transport processes in the turbulent shear flow. Accordingly, the finite-size model nanowires transferring the signals are modeled as pairs of neighbor sites $(x, y)$ that transmit a signal from $x$ to $y$ while hopping effectively to a position $(x', y')$ close by. These effective hoppings are realized so that information portion about orientation of two neighbor sites coupled into a pair is transferred from the position $(x, y)$ to $(x', y')$ (Fig. 1). Generally, each pair of neighbor sites can be considered coinciding with both ends of a model nanowire.

The pattern representing state of the $A_i(T)$ is composed of sites covered at the stage $T$ only. In course of the REP, each site element being active at a stage $T$ sends unconditionally the identical messenger particles, represented by a signal $C$, to all its nearest neighbors at the step of evolution, from $T$ to $T+1$, and thus loses its activity. One assumes that the messenger particle $C$ coming to a site element (either activated or not) from a nearest neighbor one can be effectively redirected to any site element being the nearest neighbor of this intermediate one and thus the signal $C$ is being enhanced to the $C_r$ that is already able to activate a nanowire at its sending end (Fig. 1). A nanowire activated can act as a carrier hopping to a position close by while transmitting certain signal $Q$ emitted from its sending end towards the receiving one and preserving its orientation in the discrete space (directions of the nanowire displacements agree with direction of transmitting the $Q$). The $Q$ itself is *not* the signal activating a site element, transmission of the $Q$ by the nanowire is required to change binding properties at the nanowire ends: Then the sending end can recognize a site element that was active at a stage $T$ and this, if successful, allows reception of the nanowire at its receiving end by the inactive site element occupying the same site as this end at the stage $T+1$. This reception makes this site element active at the stage $T+1$; let us call it *target site element* occupying the *target site*. This reception of at least one active nanowire is the only way of activating a site element in the REP. The scenario of the local information transfer just presented is specific to the **indirect transfer process - ITP**. Its particular case, in which the only redirection of the $C$ is back to the site element sending it, is called the **direct transfer process - DTP**, thus DTP contributes to the ITP. Specific features of the CE accomplishment represented by the $\int F(T)$, which are attributed to transferring the local information by the hopping oriented pairs (finite-size nanowires), are indicated as **finite-size effects, F-SE**. Note that effects attributed just to finite sizes of the subsystem spatial domains are called here **finiteness effects** and are relevant to projecting regularity of the hexagonal order of arrangement of the site elements onto the CE (Sec. 2.3).

### 2.3 Model Parameters

Note that nanowires with coinciding sending ends at all the six initial positions could experience hopping after being activated by one messenger particle $C_r$ if no ambient constrains were imposed (Fig. 1). Diminishing of this number can result from imposing an effectively directed forcing being the same, synchronously, within all spatial domains of the ensemble subsystems. This effect of the forcing is modeled in the same way in all the realizations $A_i$: by the same specific assignment of the hopping probabilities $P_j$ to all pairs $(x, y_j)$ within the neighborhood $S_x$ for all sites $x = x_k$ with $k \leq N$, constituting the spatial domain (see Fig. 2a), $P_1 = P_6 = P_7 \equiv 0$, $P := P_2 = P_3 = P_4 = P_5 \leq 0.25$, and, with the purpose to obtain the global direction of forcing, by initiating the expansion from the same pair of sufficiently long vertical chains (we identify them as the initiating structure $IS$, Fig. 2b; here, one considers two chains for the reliability purpose only). Eventually, the model condition of hopping requires satisfying the inequality, $P_x(y_j) \geq r_x(y_j, i)$ with $r_x(y_j, i)$ being an element assigned to the pair $(x = x_k, y_j)$ in a realization $A_i$ from the pseudorandom matrix, $r_{x_k}(\xi_j, i)$ given by certain generator for $k \leq N$, $j \leq 6$ ($\xi_j = y_j$ for $j \leq 6$, $\xi_7 = x_k$), $i \leq M$. Eventually, the conditions assumed result in development of the areas covered within right half-plane from the IS. The REP realizations, $A_i$ can be sets of more or less dispersed rugged islands that is specific to modeling the finite-size effects; compare (Fig. 2b) the $A_i$ of REP-ITP to a pattern of a state of spreading percolation - SPP that is a process without the F-SE.

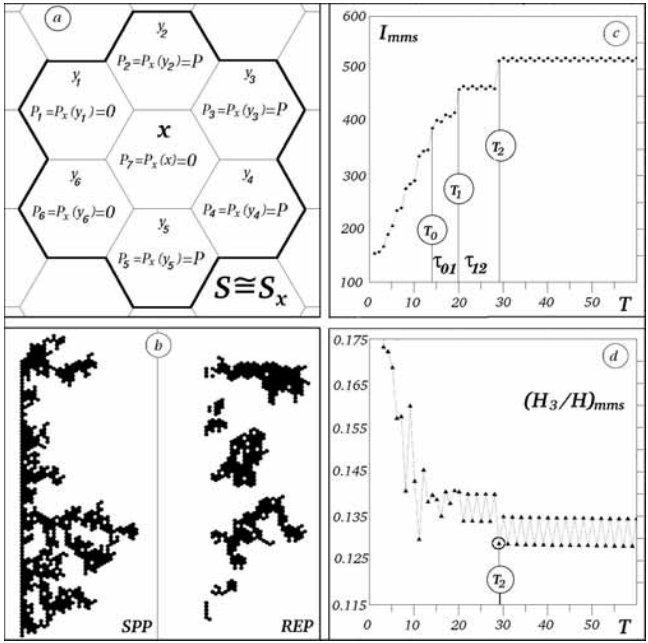

Figure 2: (a) A closest neighborhood $S_x$ of a site $x$ with assignment of the $P_j$. (b) State of the REP-ITP realization $A_i$ compared to a state of the SPP. (c,d) Characteristics of the REP advancement, $I_{mms}(T)$ (c) and structural bi-state stabilization in the MMS (d), the same value $T_2$ is identified in all graphs $(H_n/H)_{mms}(T)$ with $2 \leq n \leq 7$; REP-ITP, $\alpha = 1/3$, $M=M_0=100$, P=0.1311.

Generating the pseudorandom matrix noted above with the $M \geq M_0 \approx 100$, where $M_0$ limits from below the number of realizations being moderate, and with a number $N \geq N_0 \approx 18000$ of sites constituting the space of each component subsystem of the ensemble has appeared sufficient to obtain the required spatial-specific and ensemble-specific diversity of the REP. We have found that changes in number $N$ only, $N_0 \leq N \leq 16N_0$ do not affect the simulation results remarkably. Stepwise growth in the number I.mms(T) of covered sites constituting the sets $\int F(T)$ (see Fig. 2c) can be attributed to the discreteness of the ensemble only and the steps become less pronounced with the $M$ growing to sufficiently high values. The specific form of increments, $\delta \int F_{T_j}$ in the MMS at the characteristic steps $T_j$, which is the dispersion of the increment throughout the **entire** MMS pattern, (see Sec. 2.4, right pattern in Fig. 3), is one of manifestations of the F-SE which is relevant to opportunity for employing the REP simulations to do feasibility research. For the REP-ITP, decaying this form with growing $M$ to the increment constituted only by the sites occurring within leading part of the MMS pattern has been accepted as criterion being used to assess upper limit, $M_1$ to the values $M$ being moderate (here, for the REP-ITP, we have got $260 < M_1 < 280$). For REP-DTP, the noted change in form of the increment $\delta \int F_{T_j}$ has not been observed even with very high values of the $M$.

The finite discrete space, being here the domain of each subsystem, is an example of a compact space that can be covered by finite number of open sets having a structural property in common, which is then inherited by whole the space (e.g. [4]). Here, this is the form of a regular hexagon. Extensive discussion of modelling the **lattice order projecting efficiency - LOPE** has been presented by us previously [5] for hexagonal and squared tiling. The only method allowing *smooth* regulation of the LOPE is used here: One selects randomly a fraction $\alpha$ of sites, for which covering all their six nearest neighbors is allowed; for each site $x$ from half of the rest, i.e. $(1-\alpha)/2$ of sites, covering only odd sites within its closest vicinity $V_x$ is allowed, whereas only even sites in $V_x$ can be covered for each remained $x$ (see Fig. 2a). Thus $0 \leq \alpha \leq 1$, parameterizes the LOPE; modeling the LOPE for REP requires extention of understanding the action of *covering* the neighbor sites to more general action that is *communicating to* neighbor sites.

## 2.4 Results and Inferences

One may imagine free gliding window $S$ congruent with closest neighborhood $S_x$ of a site $x$ (Fig. 2a) and consider the $S$ as an information channel; this is in accordance with the approach known from the information theory [6]. Generally, a pattern that can be observed within the $S$ can be considered as the information carrier passing through this channel. Accordingly, distinct clusters composed of certain number $n$ of sites ($2 \leq n \leq 7$) are independently transfered by the carrier through this channel while the free window $S$ appears once at each site within the domain considered. We have defined certain characteristic $H_n$ expressing the contribution from passing the n-site clusters through the channel $S$ to mean expected amount of information that can be transferred by the carrier through the $S$; the sum, $H=\sum_{n=2}^{7} H_n$ is used here as reference scale for the $H_n$ (Fig. 2d). Distributions, $(H_n/H)_{mms}$ versus $T$ reveal appearance of a structural bi-state recognized in the sequence $\int F(T)$ of the MMS patterns in course of the REP evolution. Then, a pair of neighbor stages computed, $(T, T+1)$ corresponds to a single physical stage when the system may appear in one or in the other structural state as well. This is manifestation of the finite-size effects (we have reported previously [3] result of this type as attributed to the F-SE also and then its relevance to experiments has been noted). Contribution from REP-DTP to REP-ITP is required for bi-state stabilization in the REP-ITP (Sec. 2.2). In course of the REP development, the $T_2$ is such first instant of a jump that occurs between stabilized bi-states as observed in all the distributions $(H_n/H)_{mms}$ with $2 \leq n \leq 7$; the accumulation periods, $\tau_{12}$, $\tau_{01}$ separate the two instants, $T_1$, $T_0$ of the preceding jumps (Fig. 2c,d). The distributions of these instants, $T_j$ of the bi-state stabilization events versus the model forcing factor, $P$ (Sec. 2.3) reveal quasiperiodic character within the whole interval of LOPE parameter, $0 \leq \alpha \leq 1$; however, for REP-DTP, the distributions $T_j(P)$ are much less regular for $M \approx M_0$ and much more complex for high values of $M$.

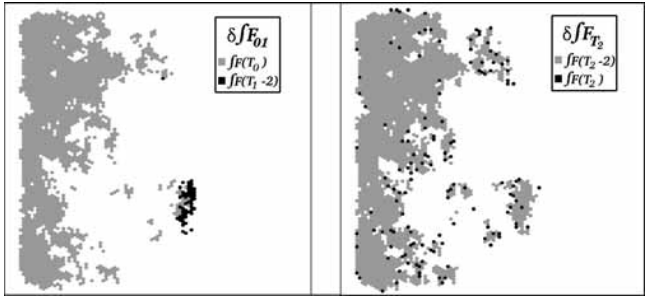

Figure 3: Pairs of MMS patterns imposed one on the other to show example of increments, for the accumulation period (left) and at the jump (right); bright pattern is imposed on the dark one; REP-ITP, $\alpha = 1/3$, $M=M_0=100$, P=0.18 whence, $T_0=82$, $T_1=92$, $T_2=101$.

Series of the MMS patterns $\int F(T)$ represents accomplishing the model collective effect in course of the REP. The $\int F(T)$, computed with $M_0 \leq M < M_1$, reveals sequence of bursts $\delta \int F_{Tj}$ in series of patterns of the MMS increments for subsequent stages of the evolution; sites representing this bursting increment are scattered throughout whole the $\int F(T_j)$ (see Fig. 3). This specific feature enables us distinguishing the $\delta \int F_{Tj}$ from neighbor increments for the accumulation periods by observing only the MMS increment patterns. Either the form of the $\delta \int F_{Tj}$ and occurrence of the nonnull increment in the MMS for an accumulation period result specifically from modeling the F-SE; increments in MMS for each of the accumulation periods following the first one with no increment in the MMS are nullsets also. In certain conditions parameterized by, $\{P, \alpha, M\}$, a burst $\delta \int F_{Tj}$ separating an accumulation period with the sedimentation type increment in the $\int F$ from the following one with no or vanishing increment in the $\int F$ enables us to detect the $T_1$ (see [1] for examples). This and the regularity in, $T_j(P, \alpha)|_{M:M_0 \leq M < M_1}$ with j=0, 1, 2, discovered for the REP-ITP suggests designing iterative procedure associating $\{P, \alpha, M\}$ of REP-ITP, representing adequately a process underlying evolution of certain real system ensemble, with the real parameters to get the real CE accomplishment with features corresponding to the specific features of model MMS growth. The noted increments in the MMS refer to the same state of the bi-state and its identification in a real system may be difficult, particularly in situation when differences between states of the bi-state, $\Delta \int F(T, T+1)$ are comparable to the increment, $\delta \int F_{Tj}$ in MMS at the jump. However, in case of the jumps between the **stabilized** bi-states, the set of sites covered which represents the difference, $\Delta \int F(T, T+1)$ is distanced from the front region of the expanding MMS whereas the increment, $\delta \int F_{Tj}$ consists of covered sites scattered throughout whole the MMS. This difference between the sets noted has been not observed for REP simulated in the squared array.

Eventually, we conclude that real processes with the scenario of local information transfer corresponding to the REP-ITP developing in the domain represented by the regular hexagonal array and parameterized by the $\{P, \alpha, M\}$ can be candidates for the regulation; then, a number of experiments required could be reduced.

## 3  PROSPECT OF FEASIBILITY

The real receiver system (**RS**) would have to appear over each of the $M$ real component systems (**CS**) for a period identified as a real effective stage $T$ of evolution. One may note two general real ways in which RS can collect actions from all the CS in accordance with the REP assumptions (Sec. 2.1).

One way allows for each CS surface projecting its action only directly towards the close congruent and opposite to it RS surface. The need to make the effect of collecting these actions independent upon the order in which the RS appears over all CS would require the RS appearing the same sufficiently high number of times over each of the $M$ component systems and each time in different order. This way would require precise displacements of the RS relatively to CS which could be realized by using commercially available solid-state linear positioning devices. This system set up would be rather complex and the requirement of collecting the actions as the aperiodic sequences of frames could make finding correspondence between real and model system parameters (Sec. 2.4) difficult.

The other way would require action able to penetrate through a stack of the layers with evolving CS and without affecting them. The RS placed over the stack of all the CS would collect their actions in better accordance with the REP assumptions than in the first way and this would facilitate using the REP-ITP simulations to regulate the real CE accomplishment within frame of one iterative procedure. It is also conceivable that the action mentioned would play role of a carrier gathering information from the subsequent CS layers to deliver it to the RS placed over the CS stack. Recent available results of research on behavior and properties of magnetic vortices and effects of breaking coherence in high temperature superconductors, particularly cuprates $YBa_2Cu_3O_{7-x}$, being affected by ambient magnetic field [7, 8] seem to provide certain suggestion about direction of search for the respective action.

## ACKNOWLEDGMENT

Parallel computations done with grant G08-16 at ICM of Warsaw Univ., support from WAR/341/202 and NATO PST.CLG.976545 is acknowledged.